\documentstyle[12pt,aps]{revtex}
\def\vec#1{{\bf #1}}
\def\mat#1{{\hat{#1}}}

\begin{document}
\sloppy
\unitlength=1mm
\begin{center}
  {\large\bf Autler -- Townes doublet probed by strong
    field}\\\vspace{0.2cm}
  {M~G~Stepanov\footnote{E-mail:
    Stepanov@iae.nsk.su}}\\\vspace{0.2cm}
  {\it Institute of Automation and Electrometry, Russian
    Academy of Sciences,\\ Siberian Branch, Novosibirsk
    630090, Russia}
\end{center}
\begin{quote}
  \hskip18pt This paper deals with the Autler -- Townes
  doublet structure. Applied driving and probing laser
  fields can have arbitrary intensities. The explanation is
  given of the broadening of doublet components with the
  growth of probing field intensity, which was observed in
  experiment. The effects of Doppler averaging are
  discussed.
\end{quote}
\begin{quote}
\end{quote}

\section{Introduction}
\label{in}

The spectroscopy of three-level systems occupies a highly
important place in nonlinear spectroscopy. In addition to
saturation or power broadening of resonance \cite{prKS48},
new remarkable effects appear: field splitting
\cite{prAT55} and truly coherent processes (such as Raman
scattering or two-photon processes). The most clear issue
is the absorption spectrum of three-level systems
interacting with two monochromatic waves within
perturbation theory in intensity of one of the waves ---
probe field spectrum
\cite{prlFJ68,prFJ69,jPPRF69,jlVCS70,zpHT70}. Some efforts
were made to construct the theory where both waves are
strong \cite{jpbS77,jPKM98}. In \cite{jpbS77}, the
coherences in the equations for the density matrix were
excluded, and equations containing only populations were
derived. As a result, the probabilities of transitions
induced by fields were renormalized and expressed in terms
of populations only. Unfortunately, our intuition in
predicting the behavior of such systems is poor yet, it is
difficult to foresee the result without solving the
equations. The computation of three-level system with a
great number of parameters (relaxation constants of
levels, waves detunings and strengths) \cite{jPKM98} is
hard to analyze.

However, power of light has increased since early
experiments, as one needs to get better light conversion
or more controllable setup. In experiment \cite{W82}
$\Lambda$-scheme interacting with two waves exhibits, at
first sight, strange behavior.  When the intensity of
comparatively weak wave increases, the components of
Autler -- Townes doublet broaden in its absorption
spectrum; finally the doublet is transformed to a single
line.

A three-level system with two strong fields can be a part
of the scheme of four-wave resonant mixing. The study of
this area is progressing rapidly, since it gives a hope to
obtain coherent CW short-wave radiation. To understand the
whole picture it is useful at first to consider the part
of a system that interacts with strong fields.

The aim of the present paper is to analyze qualitatively
the effects arising from increasing the power of weak wave
in $\Lambda$-scheme. In what follows we discuss the
general features of multi-level systems and the
possibility of reducing them to simple two-level systems.
We use the probability amplitudes approach and the
``dressed'' state concept (see \cite{CT77,praBS82}) to
understand the spectra. The energies of ``dressed'' states
are identified with the poles of the Green function of
master equations in $\omega$-representation. We are not
interested in intrinsic relaxation processes in an atomic
system and this allows us to avoid introducing the density
matrix. It reduces the system of $N^2$ equations for the
elements of the density matrix to the system for $N$
probability amplitudes (here $N = 3$).

The paper is organized as follows:

In sections \ref{be},\ref{gf} the equations of motion and
their Green function are described. In section \ref{DS}
the energies of ``dressed'' states are discussed. Section
\ref{lr} is devoted to the calculation of probe field
spectrum in a four-level system --- linear response of a
three-level one; it shows the benefits of Green function.
Section \ref{sc} presents the absorption spectra of the
strong field. In section \ref{va} we briefly consider the
question of the Doppler broadening of spectra.  In section
\ref{sm} we finally summarize essential aspects of the
paper.

\section{Basic equations}
\label{be}

Consider the atomic system consisting of three excited
levels and its resonant interaction with two
electromagnetic waves with frequencies $\omega_1$,
$\omega_2$ (see figure 1). Hereafter we assume
that the wave amplitudes are not changed essentially by
interaction (optically thin media). Thus, we do not
consider here the effects of light propagation. One can
think that there is a thin layer with absorbing atoms and
light travels trough it almost without absorption. Here we
also do not take into account the motion of atoms, i.e.,
we do not consider the effects connected with
inhomogeneous broadening of absorption spectra. Then the
time evolution of atomic wave function is described by the
Schr\"odinger equation ($\hbar = 1$)
\begin{eqnarray}
  {\rm i}\frac{{\rm d}}{{\rm d}t} |\psi\rangle = \left(
    {\mat{\cal H}}_0+{\mat{\cal H}}_{\rm int} \right)
    |\psi\rangle,
  \quad |\psi\rangle = \sum_{i=1}^3 a_i(t) |i\rangle, \\
  {\rm i}\frac{{\rm d}}{{\rm d}t} \left( \begin{array}{c}
    a_1 \\ a_2 \\ a_3 \end{array} \right) =
  \left( \begin{array}{ccc} E_1-\frac{\rm i}2\Gamma_1 & 0 &
    -G_1^* {\rm e}^{{\rm i}\omega_1 t} \\ 0 & E_2-\frac{\rm
    i}2\Gamma_2 & -G_2^* {\rm e}^{{\rm i}\omega_2 t} \\ -G_1
    {\rm e}^{-{\rm i}\omega_1 t} & -G_2 {\rm e}^{-{\rm
    i}\omega_2 t} & E_3-\frac{\rm i}2\Gamma_3 \end{array}
    \right)
  \left( \begin{array}{c} a_1 \\ a_2 \\ a_3 \end{array}
    \right).
  \label{Eqa}
\end{eqnarray}
Here ${\mat{\cal H}}_0$ is the atomic Hamiltonian,
${\mat{\cal H}}_{\rm int}$ describes the interaction with
light; $a_i$, $E_i$, $\Gamma_i$ are the probability
amplitude, the energy, and the decay rate of state
$|i\rangle$; $G_i$ is the matrix element $\langle
3|-{\mat{\cal H}}_{\rm int}|i\rangle {\rm e}^{{\rm
i}\omega_i t}$ of interaction. Equation (\ref{Eqa}) is
written in resonant (or rotating wave) approximation,
which is valid if $G_i, \omega_i-E_3+E_i \ll E_3-E_i$, $i
= 1,2$. The frequency $\Omega_i = \omega_i-E_3+E_i$ is
the detuning of the wave $i$ from resonance.

Since there is no closed cycles made up from wave fields
(see figure 1(a)) one can bring (\ref{Eqa}) to the
form without explicit dependence on time:
\begin{eqnarray}
  A_i = a_i\,{\rm e}^{{\rm i}(E_i-\Omega_i)t}, \quad i=1,2,
    \quad A_3 = a_3 {\rm e}^{{\rm i}E_3 t}, \\
  {\rm i}\frac{{\rm d}}{{\rm d}t} \left( \begin{array}{c}
    A_1 \\ A_2 \\ A_3 \end{array} \right) =
  \left( \begin{array}{ccc} \Omega_1-\frac{\rm i}2\Gamma_1
    & 0 & -G_1^* \\ 0 & \Omega_2-\frac{\rm i}2\Gamma_2 &
    -G_2^* \\ -G_1 & -G_2 & -\frac{\rm i}2\Gamma_3 \end{array}
    \right)
  \left( \begin{array}{c} A_1 \\ A_2 \\ A_3 \end{array}
    \right).
  \label{EqA}
\end{eqnarray}
Moreover, fields $G_i$ can be thought of as real.
Nevertheless, sometimes we will write $G_i^*$ instead of
$G_i$.  It helps to determine the interaction process
which corresponds to the written expression.

In spectroscopy, the observed quantities are the powers
absorbed from the waves. Light field induces transitions
between atomic levels, and absorption is connected with the
change of levels population. The time derivative of
population on level 2 is equal to
\begin{equation}
  \frac{{\rm d}}{{\rm d}t} |A_2|^2 = -\Gamma_2 |A_2|^2-
    {\rm i} G_2 A_2 A_3^* + {\rm i} G_2^* A_2^* A_3.
  \label{DtA2}
\end{equation}
The first term in r.h.s. of (\ref{DtA2}) is due to the
decay of the state $|2\rangle$. Other two terms are
responsible for the absorption of the 2nd wave.

The formulation of the problem of light absorption is the
following: at the moment $t_0$ the atom is excited to the
state $\vec A = {\vec A}_0$. Then the time evolution of
$\vec A$ is governed by (\ref{EqA}), and the power
absorbed from the 2nd wave by this atom is equal to
\begin{equation}
  P({\vec A}_0, \Omega_2, t_0) = \omega_2 \int_{t_0}^\infty
    {\rm d}t \, \left( {\rm i} G_2 A_2 A_3^* - {\rm i}
    G_2^* A_2^* A_3 \right).
\end{equation}
Because equation (\ref{EqA}) is time-independent, the
power $P$ does not depend on $t_0$. Then one can set $t_0
= 0$.  In addition, $P$ is a quadratic functional of
${\vec A}(t)$, and equation (\ref{EqA}) is linear. Thus,
$P$ can be written as $P({\vec A}_0, \Omega_2) = \omega_2
{\vec A}_0^\dagger {\mat{\cal P}}(\Omega_2) {\vec A}_0$,
where $\cdot^\dagger$ is the Hermitian conjugation. The
matrix ${\mat{\cal P}}$ contains complete information
about the absorption of the 2nd wave, i.e., using
${\mat{\cal P}}$ one can calculate $P$ with any initial
excitation ${\vec A}_0$.  For example, ${\cal P}_{22}$ is
equal to the probability of absorption of a photon from
the 2nd wave if the atom was initially excited to level 2.
Because $P$ is real, the matrix ${\mat{\cal P}}$ is
Hermitian.

\section{Green function}
\label{gf}

Let us consider equation (\ref{EqA}) with the initial
condition ${\vec A}(t = 0) = {\vec A}_0$:
\begin{equation}
  {\rm i}\frac{{\rm d}}{{\rm d}t} {\vec A} = {\mat H} \vec
    A + {\rm i} \delta(t) {\vec A}_0, \quad
  {\vec A}(t) = {\vec 0} \quad {\rm if}\,\,\, t<0.
\end{equation}
Since this equation is time-independent, it is convenient
to consider its Fourier transform, i.e, its energy
representation ($\hbar = 1$)
\begin{equation}
  {\vec A}(t) = \int \frac{{\rm d}\omega}{2\pi} \, {\rm
    e}^{-{\rm i}\omega t} {\vec A}_\omega, \quad
  \omega {\vec A}_\omega = {\mat H} {\vec A}_\omega + {\rm
    i} {\vec A}_0.
\end{equation}
It is a set of linear algebraic equations. Its solution is
equal to
\begin{equation}
  {\vec A}_\omega = {\rm i} \big( \omega-{\mat H}
    \big)^{-1} {\vec A}_0 = {\rm i} {\mat{\cal D}}(\omega)
    {\vec A}_0.
\end{equation}
The matrix ${\mat{\cal D}}(\omega)$ is the Green function
of (\ref{EqA}) in $\omega$-representation; it is also the
resolvent of the Hamiltonian $\mat H$. The matrix element
${\cal D}_{ij}(\omega)$ describes transitions $i
\leftarrow j$ induced by fields. As a function of complex
variable $\omega$, the matrix ${\mat{\cal D}}(\omega)$ has
poles at the eigenvalues $\lambda_I$, $I=1,2,3$ of the
Hamiltonian $\mat H$, i.e., at the quasi-energies of the
stationary states of ``atom+field'' system (``dressed''
states).  The energy $\lambda$ of a ``dressed'' state is
found from the equation ${\rm Det\,}(\lambda-{\mat H}) =
0$:
\begin{eqnarray}
  \left( \lambda-\Omega_1+{\textstyle\frac{\rm
    i}2}\Gamma_1 \right)
  \left( \lambda-\Omega_2+{\textstyle\frac{\rm
    i}2}\Gamma_2 \right)
  \left( \lambda+{\textstyle\frac{\rm i}2}\Gamma_3
    \right) \nonumber \\
  = \left( \lambda-\Omega_1+{\textstyle\frac{\rm
    i}2}\Gamma_1 \right)|G_2|^2+
  \left( \lambda-\Omega_2+{\textstyle\frac{\rm
    i}2}\Gamma_2 \right)|G_1|^2.
  \label{Det}
\end{eqnarray}
Each $\lambda_I$, $I=1,2,3$ has negative imaginary part
${\rm Im\,}\lambda_I<0$ corresponding to the decay rate of
the ``dressed'' state $|I\rangle\!\rangle$. In other
words, ${\mat{\cal D}}(\omega)$ has poles only in the
lower half-plane.

If the fields $G_1$, $G_2$ are weak, then there is a hope
that the evolution of the system is little different from
the evolution without fields. One can use this ``bare''
evolution as a starting point and expand the Green
function ${\mat{\cal D}}(G_1,G_2)$ into a power series in
the neighbourhood of $G_1 = 0$, $G_2 = 0$. We denote
\begin{eqnarray}
  {\mat E} = \left( \begin{array}{ccc} \Omega_1-\frac{\rm
    i}2\Gamma_1 & 0 & 0 \\ 0 & \Omega_2-\frac{\rm
    i}2\Gamma_2 & 0 \\ 0 & 0 & -\frac{\rm i}2\Gamma_3
    \end{array} \right), \quad
  {\mat G} = -\left( \begin{array}{ccc} 0 & 0 & G_1^* \\
    0 & 0 & G_2^* \\ G_1 & G_2 & 0 \end{array} \right).
\end{eqnarray}
Then ${\mat H} = {\mat E}+{\mat G}$. One can write the
following expansion:
\begin{equation}
  {\mat{\cal D}} = {\mat D} + {\mat D} {\mat G} {\mat D} +
    {\mat D} {\mat G} {\mat D} {\mat G} {\mat D} + \dots =
    {\mat D} + {\mat D} {\mat G} {\mat{\cal D}}
  \label{G},
\end{equation}
where the argument of ${\mat{\cal D}}$ and ${\mat D}$ is
equal to $\omega$, and ${\mat D}(\omega) = (\omega-{\mat
E})^{-1}$.  This series is an analog of $(1-x)^{-1} = 1 +
x + x^2 + \dots$. ``Bare'' Green function ${\mat
D}(\omega)$ has poles at the energies of ``bare'' states
$\Omega_1-\frac{\rm i}2\Gamma_1$, $\Omega_2-\frac{\rm
i}2\Gamma_2$, $-\frac{\rm i}2\Gamma_3$. The summation
(\ref{G}) shifts the poles.

The diagonal element ${\cal D}_{ii}$ is the Green function
of atom in the state $|i\rangle$. It also has poles at
$\lambda_I$. There is an admixture of all dressed states
$|I\rangle\!\rangle$ in $|i\rangle$. In the limit of weak
fields ${\cal D}_{ii}$ has obvious pole. If the fields
induce transitions between the states $|i\rangle$ and
$|j\rangle$, then the atom in the state $|i\rangle$ lives
in the state $|j\rangle$ for some time. Then the pole
corresponding to $|j\rangle$ appears in its Green
function. When the fields become stronger these poles
shift.

Now we can write a simple expression for the matrix
${\mat{\cal P}}(\Omega_2)$ describing the absorption of the
2nd wave.
\begin{equation}
  P({\vec A}_0, \Omega_2) = \omega_2 \int_0^\infty {\rm d}t
    \, {\vec A}^\dagger(t) {\mat P} {\vec A}(t), \quad
  {\rm i} {\mat P} = \left( \begin{array}{ccc} 0 & 0 & 0 \\
    0 & 0 & G_2^* \\ 0 & -G_2 & 0 \end{array} \right).
\end{equation}
Vector ${\vec A}(t)$ can be rewritten in terms of the Green
function. We obtain
\begin{eqnarray}
  {\mat{\cal P}}(\Omega_2) = \int_0^\infty {\rm d}t
    \int \frac{{\rm d}\omega_1 {\rm d}\omega_2}{(2\pi)^2}
    \, {\rm e}^{{\rm i}(\omega_1-\omega_2)t} {\mat{\cal
    D}}^\dagger (\omega_1) {\mat P} {\mat{\cal
    D}}(\omega_2) \nonumber \\
  = {\rm i} \int \frac{{\rm d}\omega_1{\rm
    d}\omega_2}{(2\pi)^2} \, \frac{{\mat{\cal D}}^\dagger
    (\omega_1) {\mat P} {\mat{\cal D}}(\omega_2)}
    {\omega_2-\omega_1-{\rm i}0} = \int \frac{{\rm
    d}\omega}{2\pi} {\mat{\cal D}}^\dagger (\omega) {\mat
    P} {\mat{\cal D}}(\omega).
  \label{EqDPD}
\end{eqnarray}
The correction ${\rm i}0$ in (\ref{EqDPD}) is due to
causality. The physical determination of the integral over
$t$ consists in multiplying by a factor ${\rm
e}^{-\varepsilon t}$ with $\varepsilon\rightarrow +0$.
Since $\omega$ (as well as $\Omega_1$ and $\Omega_2$) is
real on the contour of integration, in the integral we can
replace ${\mat{\cal D}}^\dagger (\omega)$ by ${\mat{\cal
D}}^c (\omega)$. The operation $\cdot^c$ is defined as
follows: $\cdot^c \stackrel{\rm def}= \cdot^{\dagger}$,
but during this operation the frequencies $\omega$,
$\Omega_i$, $i=1,2$ are thought to be real, i.e., they
need not be complex conjugated. The operation $\cdot^c$
changes only the sign of $\Gamma$s. Thereafter the
integrand in (\ref{EqDPD}) is an analytic function of
$\omega$, $\Omega_i$, i.e., it does not depend on
$\omega^*$, $\Omega_i^*$. Moreover, it is rational; one
can apply residue theory to calculate the integral. In its
turn, ${\mat{\cal P}}(\Omega_2)$ is an analytic function
of $\Omega_2$ and the poles of ${\mat{\cal P}}$ correspond
to the resonances in the 2nd wave absorption spectrum. The
real part of the pole position gives the frequency of the
resonance, the imaginary part represents the resonance
width.

The matrix ${\hat H}$ has right and left eigenvectors:
${\hat H} |I\rangle\!\rangle = \lambda_I
|I\rangle\!\rangle$, $\langle\!\langle I| {\hat H} =
\langle\!\langle I| \lambda_I$. They can be normalized so
that $\langle\!\langle I| J\rangle\!\rangle =
\delta_{IJ}$. Then
\begin{equation}
  {\hat H} = \sum_{I=1}^3 |I\rangle\!\rangle \lambda_I
    \langle\!\langle I|, \quad {\mat{\cal D}}(\omega) =
    \sum_{I=1}^3 |I\rangle\!\rangle
    \frac1{\omega-\lambda_I} \langle\!\langle I|.
  \label{braket}
\end{equation}
If ${\mat{\cal P}}(\Omega_2)$ has a pole at $\Omega_2 =
\Omega$, then the integral (\ref{EqDPD}) for $\Omega_2 =
\Omega$ is divergent. The contour of integration bypasses
the poles of ${\mat{\cal D}}(\omega)$ and ${\mat{\cal
D}}^c(\omega)$ in different directions. Then the integral
(\ref{EqDPD}) may be divergent only if ${\mat{\cal D}}$
and ${\mat{\cal D}}^c$ have at least one common pole
$\lambda_I = \lambda_J^c$. When $\lambda_I$ and
$\lambda_J^c$ merge together they squeeze the integration
contour, and the integral (\ref{EqDPD}) acquires
singularity. Time evolution of the dressed state
$|I\rangle\!\rangle$ is described by the exponent ${\rm
e}^{-{\rm i}\lambda_I t}$.  The case where $\lambda_I$ is
close to $\lambda_J^c$ corresponds to the resonance
between the waves and the dressed states
$|I\rangle\!\rangle$, $|J\rangle\!\rangle$.  The
absorption matrix ${\mat{\cal P}}(\Omega_2)$ contains
resonance denominators of the form $\lambda_I -
\lambda_J^c$.

\section{Dressed states}
\label{DS}

Here, for simplicity, we neglect relaxation constants. In
fact, the resonance widths under power broadening depend
on the ratio of $\Gamma$s even when the latter are small.
Therefore, the contents of this paragraph is useful for
the case of equal relaxation constants (or, maybe, when
they are comparable to each other) or for finding
resonance positions in the absorption spectrum of
auxiliary probe field. The energies of the dressed states
satisfy algebraic equation (\ref{Det}).  Let us write it
in the form
\begin{equation}
  \big[ \lambda(\lambda-\Omega_1)-|G_1|^2 \big]
    (\lambda-\Omega_2) = (\lambda-\Omega_1) |G_2|^2.
  \label{Det2}
\end{equation}
If we consider the energy levels of the system
``atom+field'', then we will see the triples of close
levels. For weak fields these levels are:
$|3\rangle|n_1,n_2\rangle$ (atomic state $|3\rangle$,
$n_1$ quanta in the 1st field, and $n_2$ quanta in the 2nd
one), $|1\rangle|n_1+1,n_2\rangle$, and
$|2\rangle|n_1,n_2+1\rangle$. If we consider the system
within perturbation theory, we use the atomic basis of
states. They are close in energy; in resonant
approximation the field-induced transitions occur only
inside each triple. The ``dressed'' energy levels are
found from the secular equation for each triple
(\ref{Det}).  In resonant approximation the triples are
independent.

The resonances in the spectra occur when two eigenvalues
$\lambda$ are close to each other. The minimal distance
between them gives the order of resonance width. The
general idea is that one can consider two close $\lambda$s
separately from others and write the quadratic secular
equation for them. These two eigenvalues are pushed apart
by the intermixing --- the general scenario due to the
Hermitity of interaction.

When $G_2 \ll G_1, \Omega_{1,2}$, $\lambda$ is found from
the condition of vanishing of l.h.s. in (\ref{Det2}):
\begin{equation}
  \lambda_{1,2} = \Big( \, \Omega_1 \pm \big( \Omega_1^2 +
    4|G_1|^2 \big)^{1/2} \, \Big)/2, \quad \lambda_3 = \Omega_2.
  \label{l123}
\end{equation}
First two dressed states result from $|1\rangle$ and
$|3\rangle$ by the 1st wave field splitting. The third one
corresponds to $|2\rangle$. The 1st wave makes the
distance between $\lambda_1$ and $\lambda_2$ not less than
$2|G_1|$.  Then the resonances in the 1st wave absorption
spectrum undergo power broadening --- their width becomes
proportional to $|G_1|$.

When $\Omega_2$, i.e., $\lambda_3$, is close to
$\lambda_1$, we can consider $\lambda_1$ and $\lambda_3$
separately from $\lambda_2$. Denoting $\Omega_2 =
\lambda_1 + \Omega$, $\lambda = \lambda_1 + \Lambda$, we
obtain
\begin{equation}
  M_{1,2} = \frac12 \left( 1 \mp
    \frac{\Omega_1}{\big(\Omega_1^2 +4|G_1|^2\big)^{1/2}} \right),
    \quad \Lambda (\Lambda-\Omega) \simeq M_1 |G_2|^2.
  \label{Det21}
\end{equation}
The energies $\lambda_1$ and $\lambda_3$ are split by the
2nd wave, whose intensity is multiplied by the factor
$M_1$. The case $\lambda_3 \simeq \lambda_2$ is handled in
a similar way, one should only substitute $\lambda_2$ and
$M_2$ for $\lambda_1$ and $M_1$. The coefficients
$M_{1,2}$ were called ``memory factors'' or ``correlation
factors'' \cite{jPPRF69,RS91}. One can diagonalize the
Hamiltonian $\mat{H}$ neglecting the 2nd field $G_2$ and
turn from bare states $|1\rangle$ and $|3\rangle$ to the
states dressed by the 1st field having energies
$\lambda_{1,2}$.  Then $\sqrt{M_{1,2}}|G_2|$ are simply
the matrix elements of interaction between these dressed
states and bare state $|2\rangle$, induced by the 2nd
field. It is well known that while the 2nd wave is weak
(probe), its absorption spectrum $P(\Omega_2)$ has two
resonance lines (Autler -- Townes doublet) \cite{prAT55}.
In this case their width is determined by the relaxation
constants. From the behavior of the dressed states we
conclude that these two lines will undergo power
broadening by the 2nd field if it becomes stronger.

\section{Probe field spectrum}
\label{lr}

We consider the scheme illustrated in figure
1(b) in the case where only the 4th level is
excited (i.e., ${\vec A}_0 = (0,0,0,1)$). We assume the
fields $G_\mu$ and $G_\nu$ to be weak. Thus, only the 1st
order of perturbation theory in $G_{\mu,\nu}$ is needed.
In the main order, $A_4 = {\rm exp\,}(-\Gamma_4 t/2 + {\rm
i}(\Omega_1 + \Omega_\mu)t) \cdot \theta(t)$, where $A_4 =
a_4\,{\rm exp\,}({\rm i}(E_4 - \Omega_1 - \Omega_\mu)t)$.
The probability amplitudes $A_i$, $i = 1,2,3$ have the
form
\begin{eqnarray}
  A_i(t) = \int {\rm d}t_0 \left( G_\mu {\cal
    D}_{i1}(t-t_0) + G_\nu {\rm e}^{-{\rm i}\varepsilon
    t} {\cal D}_{i2}(t-t_0) \right) A_4(t_0) \nonumber \\
  = \int \frac{{\rm d}\omega}{2\pi} {\rm e}^{-{\rm i}
    \omega t} \left( \frac{G_\mu {\cal D}_{i1}(\omega)}
    {\Gamma_4/2 + {\rm i}(\Omega_1 + \Omega_\mu - \omega)}
    + \frac{G_\nu {\cal D}_{i2}(\omega)} {\Gamma_4/2 +
    {\rm i}(\Omega_2 + \Omega_\nu - \omega)} \right),
\end{eqnarray}
where $\varepsilon = \Omega_2 + \Omega_\nu - \Omega_1 -
\Omega_\mu$ and ${\mat{\cal D}}(t) = \int {\rm d}\omega \,
{\rm e}^{-{\rm i}\omega t} {\mat{\cal D}}(\omega) / 2\pi$.
The power absorbed from the field $G_\mu$ is equal to
\begin{eqnarray}
  P_\mu(\Omega_\mu) = 2\omega_\mu\,{\rm Re}\,{\rm i}
    G_\mu^* \int_0^\infty {\rm d}t \,  A_4^* A_1 =
    2\omega_\mu\,{\rm Re}\,{\rm i} G_\mu^* \int \frac{{\rm
    d}\omega}{2\pi} \frac1{\Gamma_4/2 - {\rm i}(\Omega_1 +
    \Omega_\mu - \omega)} \nonumber \\
  \times \left( \frac{G_\mu {\cal D}_{11}(\omega)}
    {\Gamma_4/2 + {\rm i}(\Omega_1 + \Omega_\mu - \omega)}
    + \frac{G_\nu {\cal D}_{12}(\omega)} {\Gamma_4/2 + {\rm
    i}(\Omega_2 + \Omega_\nu - \omega)} \right) \nonumber
    \\
  = 2\omega_\mu\,{\rm Re}\,{\rm i} G_\mu^* \left(
    \frac{G_\mu {\cal D}_{11}(\omega_*)}{\Gamma_4} +
    \frac{G_\nu {\cal D}_{12}(\omega_*)} {\Gamma_4 + {\rm
    i}\varepsilon} \right),
  \label{PFS}
\end{eqnarray}
where $\omega_* = \Omega_1 + \Omega_\mu + {\rm
i}\Gamma_4/2$. The expression for the absorption
$P_\mu(\Omega_\mu)$ linearly depends on the Green
functions ${\cal D}_{11}$ and ${\cal D}_{12}$, since we
considered the linear response of the three-level system.
Note that if the initial excitation goes to levels
$1,2,3$, then the result will contain the product of two
Green functions.  If one tries to calculate the power
absorbed from the field $G_\mu$ (\ref{PFS}) using
(\ref{EqDPD}), where two Green functions and the matrix
${\mat P}$ are taken for the four-level system, then one
of these Green functions (${\mat{\cal D}}$ or ${\mat{\cal
D}}^c$) will be equal to ${\cal D}_{44} \simeq D_{44}$.

The second term in brackets in (\ref{PFS}) corresponds to
parametric process. There is a closed cycle made up from
fields $\mu$-1-2-$\nu$ with the whole detuning
$\varepsilon$. This contribution to the spectrum
$P_\mu(\varepsilon)$ has the width $\Gamma_4$. However,
this is only the gain in the thin media approximation (or
the parametric instability increment). If the parametric
process leads to instability, then the emission spectrum
of thick media has different width, which can be much
smaller.

\section{Spectrum ${\mat {\cal P}}(\Omega_2)$}
\label{sc}

Here we consider the case of small equal relaxation
constants $\Gamma_1 = \Gamma_2 = \Gamma_3 \rightarrow 0$.
We focus only on the diagonal elements of the matrix
$\mat{\cal P}$, which give the absorption power when the
initial excitation goes to the real atomic levels. For
real $G_j$, $j = 1,2$, the components $\langle i|I
\rangle\!\rangle$ of the eigenvectors of $\mat H$ are also
real. From (\ref{EqDPD}) and (\ref{braket}) immediately
follows
\begin{equation}
  {\cal P}_{ii}(\Omega_2) = 2G_2 \sum_{I \ne J}
    \frac{\langle i|I \rangle\!\rangle \langle\!\langle I|2
    \rangle \langle 3|J \rangle\!\rangle \langle\!\langle
    J|i \rangle}{\lambda_J-\lambda_I}.
  \label{EqPii}
\end{equation}
As a function of $\lambda$s, (\ref{EqPii}) is a rational
expression symmetric with respect to permutations. Then it
can be easily expressed in terms of the coefficients of
(\ref{Det}), i.e., waves detunings and intensities. Note
that ${\rm Tr\,}{\mat{\cal P}}$ is equal to zero, since
$\sum_i \langle i|I \rangle\!\rangle\langle\!\langle J|i
\rangle = \delta_{IJ}$ (in the case of equal populations of
levels the interaction is absent). The power absorbed from
the 2nd wave when the excitation goes to level 2 is given
by
\begin{eqnarray}
  {\cal P}_{22} = 2 G_2^2 P/Q, \quad I_1 = G_1^2, \quad I_1
    = G_1^2, \quad I_{12} = G_1^2 + G_2^2, \label{P22} \\
  P = (4 I_{12} + \Omega_1^2) (I_{12} + (\Omega_1 -
    \Omega_2)^2) - I_1 \Omega_2^2 - I_2 \big[ 3I_{12} +
    7\Omega_1^2 -10\Omega_1\Omega_2 + 4\Omega_2^2 \big], \\
  Q = (4 I_{12} + \Omega_1^2) (I_{12} + (\Omega_1 -
    \Omega_2) \Omega_2)^2 + I_2 (\Omega_1 - \Omega_2)
    \nonumber \\ \cdot \big[ 2(\Omega_1 - 2\Omega_2)(9
    I_{12} + (\Omega_1 + \Omega_2)(2\Omega_1 - \Omega_2)) -
    27 I_2 (\Omega_1 - \Omega_2) \big].
\end{eqnarray}

Consider the simple case $\Omega_1 = 0$. One can measure
frequency in $|G_1|$ units, i.e., set $|G_1| = 1$ and
obtain
\begin{equation}
  {\cal P}_{22}(\Omega_2) = 2 G_2^2 \frac{G_2^4 + 5 G_2^2 +
    4 + 3 \Omega_2^2} {4 (G_2^2 + 1)^3 + (G_2^4 + 20 G_2^2
    - 8) \Omega_2^2 + 4 \Omega_2^4}.
\end{equation}
The poles in the spectrum are situated at the points
\begin{equation}
  \Omega_{2*} = \pm \left( 1 - \frac{5G_2^2}2 -
    \frac{G_2^4}8 \pm \frac{G_2(G_2^2-8)^{3/2}}8
    \right)^{1/2},
\end{equation}
(for the poles positions in other simple cases see
\ref{rpa}). The asymptotics for small and large $G_2$ are
the following:
\begin{equation}
  \Omega_{2*} = \pm \left\{ \begin{array}{ll} 1 \pm \sqrt{2}
    {\rm i} G_2 - G_2^2/4, & G_2 \ll 1, \\ 2{\rm i}G_2,\
    {\rm i}G_2^2/2, & G_2 \gg 1. \end{array} \right.
\end{equation}
The poles $\Omega_{2*} = \pm {\rm i}G_2^2/2$ for $G_2 \gg
1$ are inessential, since their contribution to the
spectrum is negligible. The positions of the poles
$\Omega_{2*} = \pm 2{\rm i}G_2$ correspond to the usual
power broadening by the 2nd wave or to saturation
\cite{prKS48}. The resonance positions $\Omega_{2*} \simeq
\pm 1$ for $G_2 \ll 1$ are the positions of Autler --
Townes doublet components \cite{prAT55}. The addition
$\sqrt{2}{\rm i} G_2$ is the power broadening of these
components by the 2nd field. The term $-G_2^2/4$ is the
nonlinear shift of the resonance produced by this field
(see \ref{rsa}).

When $G_2^2 = 8$ the poles in the spectrum $P(\Omega)$
merge together: $\Omega_{2*} = \pm 3^{3/2} {\rm i}$,
${\cal P}_{22} = 4(104+3\Omega_2^2)/(27+\Omega_2^2)^2$
(see figure 4 (a)). At $G_2 = \sqrt{8}$ all dressed
state energies are equal to each other:  $\lambda_{1,2,3}
= \Omega_{2*}/3 = \pm \sqrt{3}{\rm i}$. This situation
seems to be destroyed when $\Omega_1 \ne 0$ (see figure
4(b)) or relaxation constants are not negligible.

In figure 3 one can see that the power broadening
of Autler -- Townes doublet components is proportional to
the amplitude of the 2nd wave. When $\Omega_1 \ne 0$ the
broadening is asymmetric because the rates of interaction
between level 2 and the states dressed by the 1st field
are not equal to each other ($M_1 \ne M_2$). When the 2nd
field becomes stronger the resonances come closer to each
other. When $G_2 \gg G_1, \Omega_1$ the poles in the
spectrum are situated at the points
\begin{equation}
  \Omega_{2*} = \pm 2{\rm i}G_2, \quad \Omega_{2*} =
    -\frac{G_2^2}{\Omega_1 \pm 2{\rm i}}.
\end{equation}
The first pair of poles corresponds to the simple
resonance between levels 2 and 3 taking into account the
saturation caused by the 2nd field. If we decrease the
intensity of the 2nd wave, then this resonance will be
transformed to the component of Autler -- Townes doublet
$\Omega_2 = (\Omega_1 - ( \Omega_1^2 + 4|G_1|^2)^{1/2}
)/2$. The second pair of poles corresponds to the other
component of the doublet (if the 1st field is weak, then
this line is the resonance between the virtual level and
level 2, i.e., two-photon resonance). The position of this
resonance for $G_2 \gg G_1, \Omega_1$ (the real part of
the pole position) has the sign opposite to that of
$\Omega_1$. Note that in figure 3(b) the narrower
part of the spectrum goes to the left. For extremely high
values of $G_2$ the second part of the spectrum becomes
wider than the first one (the width becomes proportional
to $G_2^2$, see dots in figure 4(b)).

The fact that the width of one of the two resonances is
proportional to $G_2^2$ for $G_2 \gg G_1, \Omega_1$
corresponds to the following: the 2nd field has the
amplitude $G_2$, which is much greater than the distance
$(\Omega_1^2 + 4|G_1|^2)^{1/2}$ between the states dressed
by the 1st field. Then the 2nd field feels these two
states as only one state; it cannot see the energy
structures with resolution higher than $|G_2|$. As the
system becomes similar to an ordinary two-level system,
only one resonance with the power width $2|G_2|$ should
remain.  There are different ways to kill the 2nd
resonance: to increase its width or to decrease its
amplitude, both ways being used here.  Mathematically, the
fourth-order algebraic equation for $\Omega_{2*}$ is
transformed to the second-order algebraic equation, i.e.,
the coefficients of $\Omega_{2*}^4$ and $\Omega_{2*}^3$
are comparatively small. It is well known that the
algebraic equation with small leading coefficient
$\epsilon$ has at least one root which is large in
parameter $1/\epsilon$.

\section{Doppler broadening of spectra}
\label{va}

One can take into account the Doppler broadening of
absorption spectra by substituting $\Omega_1 - {\vec k}_1
{\vec v}$ and $\Omega_2 - {\vec k}_2 {\vec v}$ for
$\Omega_1$ and $\Omega_2$ and integrating the result
(\ref{P22}) with some distribution of particles ${\mat
\rho}({\vec v})$. Consider the case of copropagating
waves, where only a longitudinal projection $v_{||}$ of
velocity is needed. When the Doppler width is infinite,
the answer for the spectrum is the integral of the
rational function $p_4(v_{||})/p_6(v_{||})$ over velocity,
where $p_n(x)$ is the $n$th degree polynomial. Some
simplification takes place when $k_1 = k_2$ or $k_i = 0$
for some $i = 1,2$. In the general case the expressions
for velocity-averaged ${\mat {\cal P}}(\Omega_2)$ are
difficult to derive and analyze. We will discuss the
effects of the Doppler broadening only qualitatively.

When the field $G_2$ is weak and $k_2 < k_1$, the
velocity-averaged spectrum ${\mat {\cal P}}(\Omega_2)$ has
two narrow asymmetric resonance lines coming from turning
points \cite{osBLBP82} of frequency branches
\begin{equation}
  \Omega_{2*}(v_{||}) \simeq k_2 v_{||} + \Big( \Omega_1 -
    k_1 v_{||} \pm \big( (\Omega_1 - k_1 v_{||})^2 +
    4|G_1|^2 \big)^{1/2} \Big)/2. \label{va1}
\end{equation}
The turning point is the extreme point of velocity
dependence of resonance frequencies $\Omega_{2*}$, i.e.,
at the turning point $v_{\rm tp}$ we have $({\rm
d}\Omega_{2*}/{\rm d}v_{||})|_{v_{||} = v_{\rm tp}} = 0$.
The spectra of this type are well known (see, e.g.,
\cite{BC74}). For the universal shape of asymmetric line
due to the turning point see \cite{sjlSS98}. When $G_2 \ll
\Gamma$, the characteristic width of this resonance line
is of order $\Gamma$. When $G_2 \gg \Gamma$, the resonance
line width is determined by the power broadening by the
2nd wave.  Nevertheless, while $G_2 \ll G_1$ one can use
the expression (\ref{va1}) (where the influence of $G_2$
on the resonance position is neglected) for $\Omega_{2*}$.
The rough feature of the spectrum --- two narrow
asymmetric resonance lines --- remains, but the width of
these lines depends on $G_2$ (see figure 5).

When $\Omega_1$ and $G_1$ are compared to the Doppler
width, an ``isolated peak'' appears in the spectrum. Its
width depends on $\Omega_1$, $G_1$ (see
\cite{osBLBP82,josabJND85}) and is larger than $G_2$ in the
general case. The dependence of the width on $\Omega_1$ and
$G_1$ due to inhomogeneous broadening is not a subject of
the present paper, but at certain conditions the width
decreases and becomes a value of order $G_2$.

When $G_2 \sim G_1 \gg \Gamma$, the absorption spectrum
does not contain narrow lines, i.e., all resonance widths
are of order $|G_1|$ or $|G_2|$.

If $k_2 > k_1$, then the frequency branches (\ref{va1})
have no turning points, and the absorption spectrum ${\mat
{\cal P}}(\Omega_2)$ has no narrow resonance lines even
when the 2nd wave is weak. If we increase the intensity
$G_2^2$, the narrow lines will not appear.

\section{Conclusions}
\label{sm}

Let us summarize the description of the three-level system
resonantly interacting with two strong monochromatic waves.

The expression for the spectrum contains resonant
denominators $\lambda_I - \lambda_J^*$, where $\lambda_I$
is the energy of $I$th ``dressed'' state. For $N$-level
atomic system the dressed state energies satisfy the
algebraic equation of $N$th order which is the secular
equation for $N$ levels of the system ``atom+field'' that
are close in energy.  When two dressed states are close to
each other and all other states are far from them in
energy, the secular equation can be reduced to the
quadratic secular equation for these two states.  The part
of the atomic system consisting of these two dressed states
behaves as an ordinary two-level system.

If some connected component of strong fields covers the
whole atomic system with $N$ states, then all $N$
``dressed'' states will be far from each other. As a
consequence, all resonances in the spectra will be wide.
Atomic system can be covered by strong fields, but strong
fields can be disconnected (e.g., fields $1$ and $\nu$ in
figure 1(b) are strong). Then the matrix
elements between the states dressed by field $1$ and by
field $\nu$ will be small and resonances will be narrow.

When the 2nd wave is weak we can think that it couples
level 2 (see figure 1) and levels 1, 3 split by
the 1st wave. The distance between the split levels is much
greater than the amplitude of the 2nd wave; these two
transitions cannot be resonant simultaneously. Each
transition can be treated as a two-level system. The
effects of power broadening by the 2nd wave appear. When
the 2nd wave is very strong, only the two-level system
coupled by it remains and Autler -- Townes doublet becomes
blurred.

Indeed, the power width in a two-level system depends on
the ratio of relaxation constants $\Gamma$ of levels even
when the latter are small. Thus, the power width of Autler
-- Townes component, due to the intensity of the 2nd wave,
should also depend on the ratio of $\Gamma$s. The
expressions will be similar to the well-known ones for a
two-level system: one should take $\Gamma_2$ and the decay
rate of the state ``dressed'' by the 1st field.

In experiment \cite{W82} the 1st wave detuning $\Omega_1$
was equal to zero and two components had the same power
width. It is interesting to observe the asymmetric
broadening when $\Omega_1 \ne 0$.

When the 2nd wave is generated in the media,
self-consistent problem taking into account energy losses
should be solved. The doublet gives two frequencies of
generation, which merge together at a certain rate of
reflection losses.

\medskip

The author is grateful to E~V~Podivilov and S~A~Babin for
useful discussions. The present paper was partially
supported by Soros Foundation (gr. a98-674), by INTAS (gr.
96-0457) within the program of ICFPM, and by RFBR (gr.
96-02-19052).

\appendix

\section{Resonances in probe field spectrum}
\label{rsa}

Consider the $n$-level system with dressed energies
$\lambda_1, \lambda_2, \dots, \lambda_n$ and probe field
spectrum, where the probe field $G$ with detuning $\Omega$
resonantly interacts with level $|0\rangle$ and the
$n$-level system. The dressed energies $\lambda$ of the
whole system can be found from the Hamiltonian:
\begin{eqnarray}
  {\mat H} = \left( \begin{array}{ccccc} \Omega &
    \sqrt{M_1}G & \sqrt{M_2}G & \dots & \sqrt{M_n}G \\
    \sqrt{M_1}G & \lambda_1 & 0 & \dots & 0 \\ \sqrt{M_1}G
    & 0 & \lambda_2 & \dots & 0 \\ \dots & \dots & \dots &
    \dots & \dots \\ \sqrt{M_n}G & 0 & 0 & \dots &
    \lambda_n \end{array} \right), \\
  \lambda - \Omega = \left( \frac{M_1}{\lambda - \lambda_1}
    + \frac{M_2}{\lambda - \lambda_2} + \dots +
    \frac{M_n}{\lambda - \lambda_n} \right) G^2.
    \label{nDet}
\end{eqnarray}
The resonances are situated at $\Omega \simeq \lambda_i$,
$i = 1,2,\dots,n$. When $\Omega$ is close to $\lambda_1$
the equation (\ref{nDet}) has multiple root if
\begin{equation}
  \Omega = \lambda_1 \pm 2{\rm i}\sqrt{M_1}G - \left(
    \frac{M_2}{\lambda_1 - \lambda_2} + \dots +
    \frac{M_n}{\lambda_1 - \lambda_n} \right) G^2 + O(G^3).
  \label{nW}
\end{equation}
The Green function ${\cal D}_{00}(\omega)$ has the form
\begin{equation}
  {\cal D}_{00}^{-1}(\omega) = \big( (\omega - {\mat
    H})^{-1}_{00} \big)^{-1} = \omega - \Omega - \sum_i
    \frac{M_i G^2}{\omega-\lambda_i}.
  \label{D00}
\end{equation}
The mass operator $\omega - \Omega - {\cal D}_{00}^{-1}$ is
the sum of loops through levels $i = 1,2,\dots,n$.  When
$\Omega \simeq \lambda_1$ the loop $i = 1$ is much greater
than others. This loop gives the power broadening of the
resonance $\Omega \simeq \lambda_1$. Other loops
renormalize the detuning, which gives the shift of the
resonance. Note that the shift is of order $G^2/(\lambda_1
- \lambda_i)$, which is much smaller than the width of the
resonance $\sim G$. The shift of the resonance in
(\ref{nW}) can be important only when the field $G$ is not
very weak. Then it gives qualitatively correct answer. When
the probe field $G$ is weak enough, the shift can be
neglected and the resonances $\Omega \simeq \lambda_i$ live
independently.

\section{Resonance positions in spectra}
\label{rpa}

Although it is not very difficult to calculate the integral
(\ref{EqDPD}), some properties of $\mat{\cal P} (\Omega_2)$
can be elucidated by simple algebraic manipulations. It
should be noticed that the same result (also without
integration) can be obtained using the density matrix
formalism.

The resonance position $\Omega_2 = \Omega$ in the 2nd wave
absorption spectrum can be found from the condition that
one pole of ${\mat{\cal D}}$ coincides with at least one
pole of ${\mat{\cal D}}^c$. Consider the following
combination of eigenvalues:
\begin{equation}
  {\cal Z} = \prod_{I,J} (\lambda_I-\lambda_J^c) = {\rm
    Det}\,f_{\mat{H}} \big( \mat{H}^c \big) = -{\rm
    Det}\,f_{\mat{H}^c} \big( \mat{H} \big),
  \label{Z}
\end{equation}
where $f_{\mat A}(\lambda) = {\rm Det}\,(\mat{A}-\lambda)$.
The resonance occurs when ${\cal Z} = 0$.  As a function of
$\lambda$s and $\lambda^c$s, ${\cal Z}$ is symmetric with
respect to permutations, so it can be easily expressed in
terms of detunings, fields, and relaxation constants. In
the general case the expression for ${\cal Z}$ is bulky,
hence we will consider some particular cases. We assume
that $\Gamma_1 = \Gamma_2 = \Gamma_3 = \Gamma$; $\Omega_1 =
\Omega_2 = \Omega$ (\ref{w12e0}), $\Omega_1 = 0$
(\ref{w1e0}), $\Omega_1 = \Gamma = 0$ (\ref{w1ge0}).

\begin{eqnarray}
  {\cal Z} = -\Gamma^3 Z, \quad I_1 = G_1^2, \quad I_1 =
    G_1^2, \quad I_{12} = G_1^2+G_2^2, \quad \\
  Z = \left( \Omega^2 + 4I_{12} + \Gamma^2 \right) \left(
    \Gamma^2 \Omega^2 + (I_{12} + \Gamma^2)^2 \right),
    \label{w12e0} \\
  Z = (4I_{12} + \Gamma^2) \big( (\Omega_2^2 - I_{12} -
    \Gamma^2)^2 + 4\Gamma^2 \Omega_2^2 \big) -
    I_2 \Omega_2^2 (4\Omega_2^2 - 36I_1 - 9I_2),
    \label{w1e0} \\
  Z = 4I_{12} (\Omega_2^2 - I_{12})^2 - I_2 \Omega_2^2
    (4\Omega_2^2 - 36I_1 - 9I_2). \label{w1ge0}
\end{eqnarray}

\vspace{0.5cm}\centerline{\bf FIGURE CAPTIONS}\vspace{0.2cm}

FIG.~1. 3-Level system interacting with two
  electromagnetic waves (a) and its testing by two weak
  (probe) waves (b).

FIG.~2. Dressed state energies $\lambda(\Omega_2)$, $G_1 =
  1$. One can see the asymptotics $\lambda_{1,2}$,
  $\lambda_3$ (\protect\ref{l123}). When $\lambda_3 \simeq
  \lambda_{1,2}$ the energies $\lambda$ repel each other
  with the rate $(M_{1,2})^{1/2}|G_2|$ (see
  (\protect\ref{Det21})). Figure 2(a):  $\Omega_1 = 0$,
  $G_2^2 = 1/32$ (solid curve), $G_2^2 = 1/2$ (dashed
  curve). Figure 2(b):  $\Omega_1 = 4$, $G_2^2 = 1/8$. When
  $\Omega_1 \ne 0$ we have $M_1 \ne M_2$, the splitting by
  the 2nd wave is assymmetric.

FIG.~3. Absorption power ${\cal P}_{22}(\Omega_2, G_2)$ at
  $G_1 = 1$, $\Omega_1 = 0$ (a) and $\Omega_1 = 4$ (b). The
  amplitude of the 2nd field $G_2$ varies from $0.2$ to $1.4$
  with the step $0.2$ (a), and from $0.5$ to $3$ with the
  step $0.5$ (b).

FIG.~4. Pole positions $\Omega_{2*} = ({\rm
  Re\,}\Omega_{2*}, {\rm Im\,}\Omega_{2*})$ for different
  values of $G_2$, $G_1 = 1$, $\Omega_1 = 0$ (a) and
  $\Omega_1 = 4$ (b). The dots in figure 4(a) correspond to
  $G_2 = \sqrt{n}$, $n = 0,1,\dots,8$; in 4(b) --- to $G_2
  = 0,1,\dots,5$.

FIG.~5. Velocity-averaged spectrum ${\cal
  P}_{22}(\Omega_2)$ (numerical calculations), $G_1 = 1$,
  $\Omega_1 = 0$, $k_1 = 1$, $k_2 = 0.8$, $G_2$ varies
  from $0.2$ to $1.4$ with the step $0.2$. Uniform
  velocity distribution or infinite Doppler width.

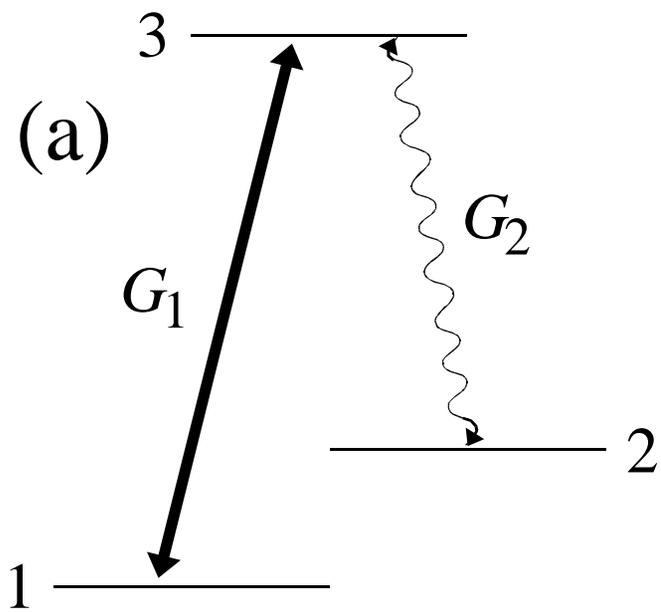
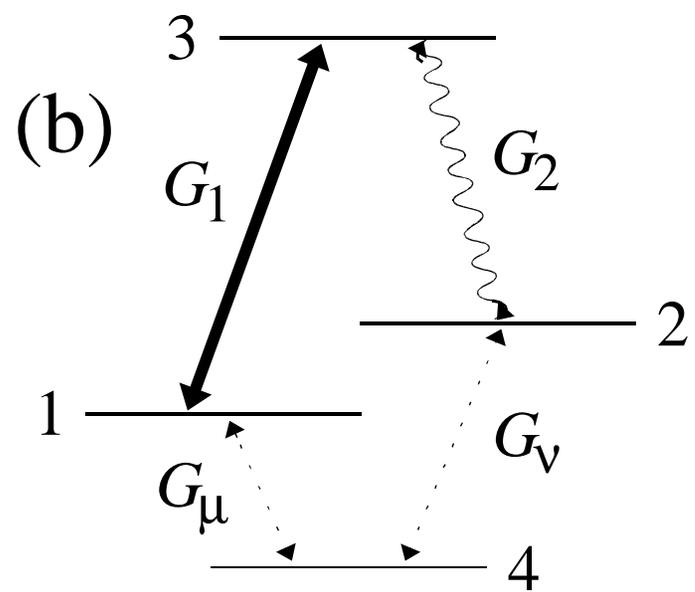
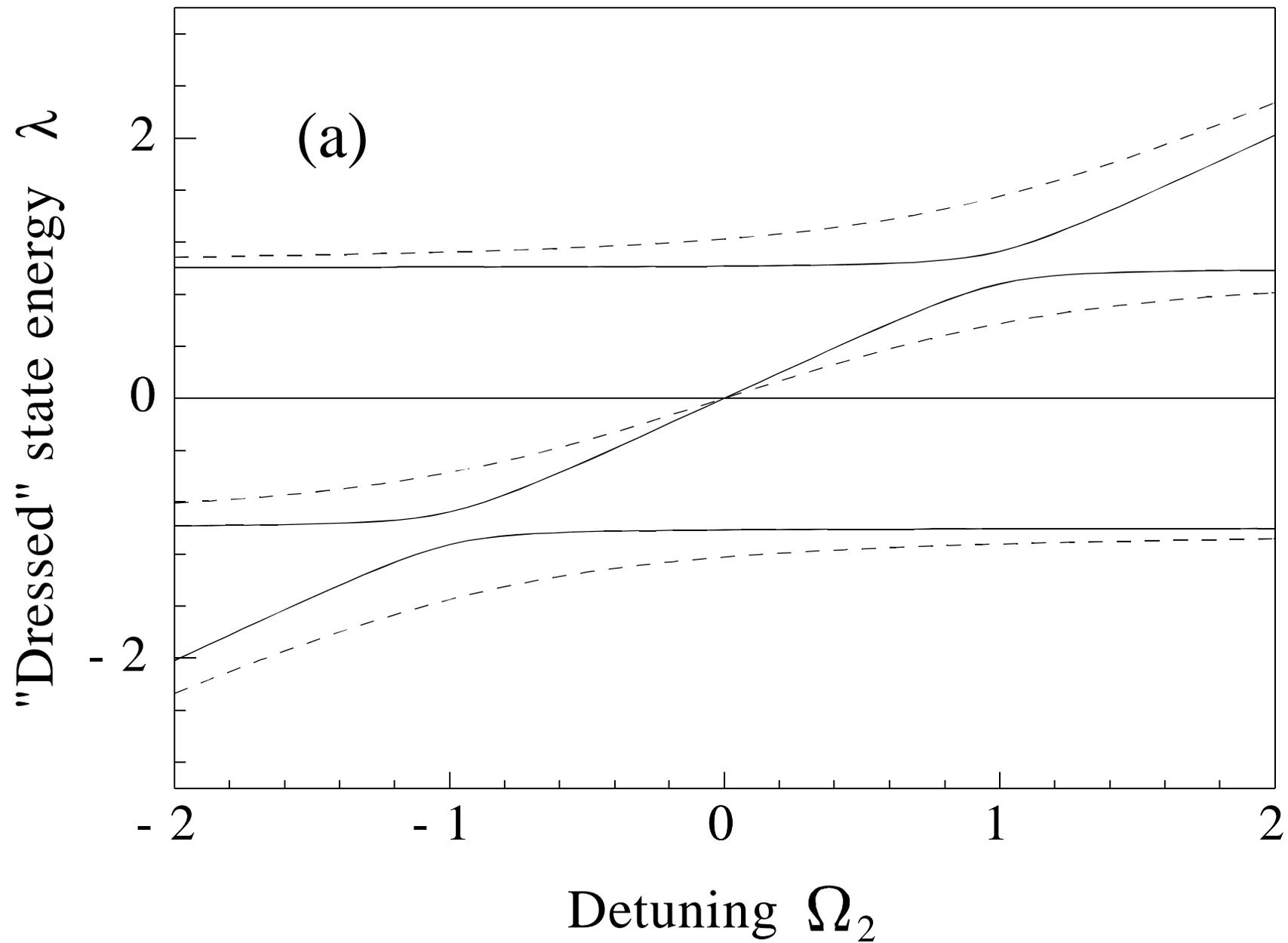
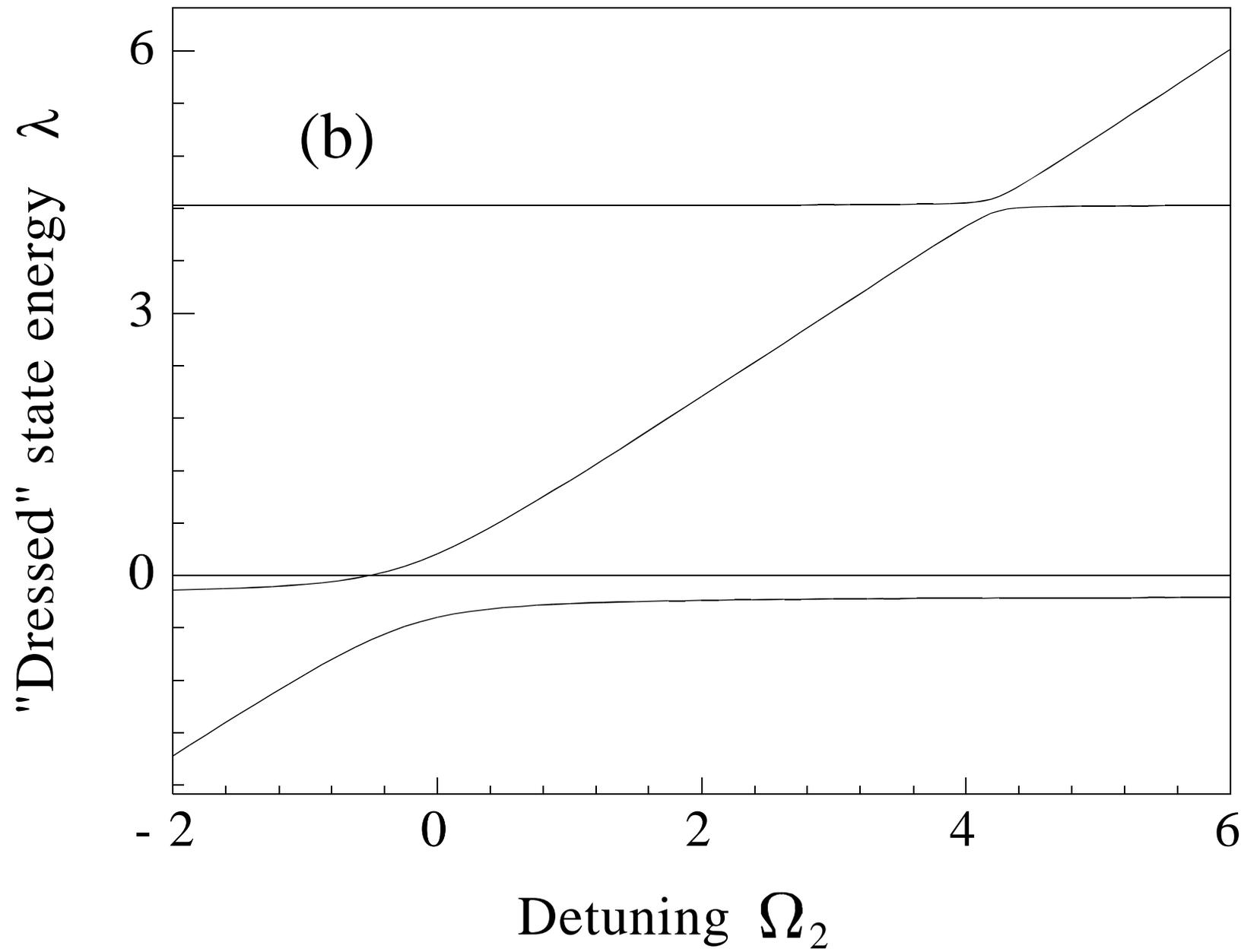
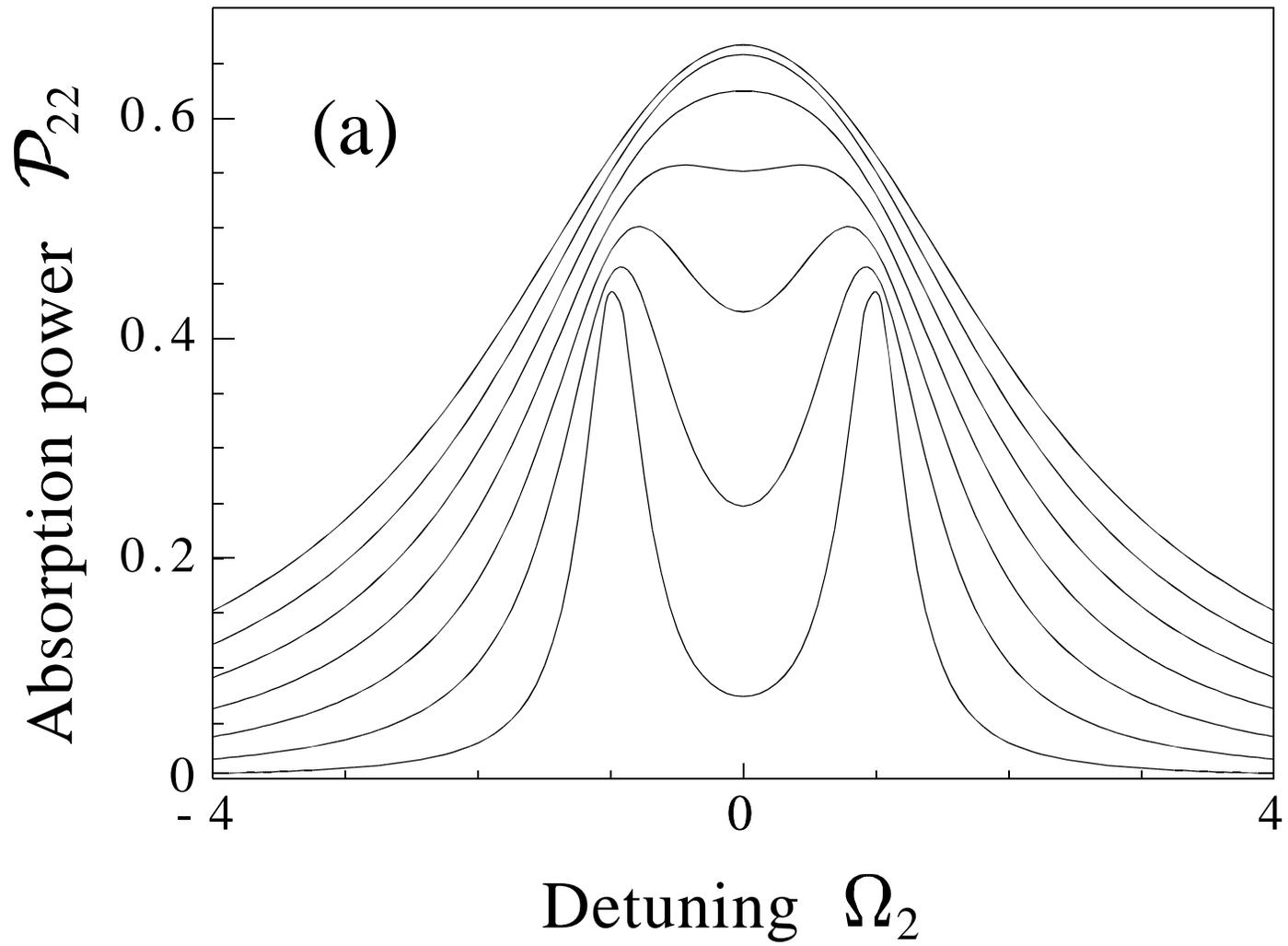
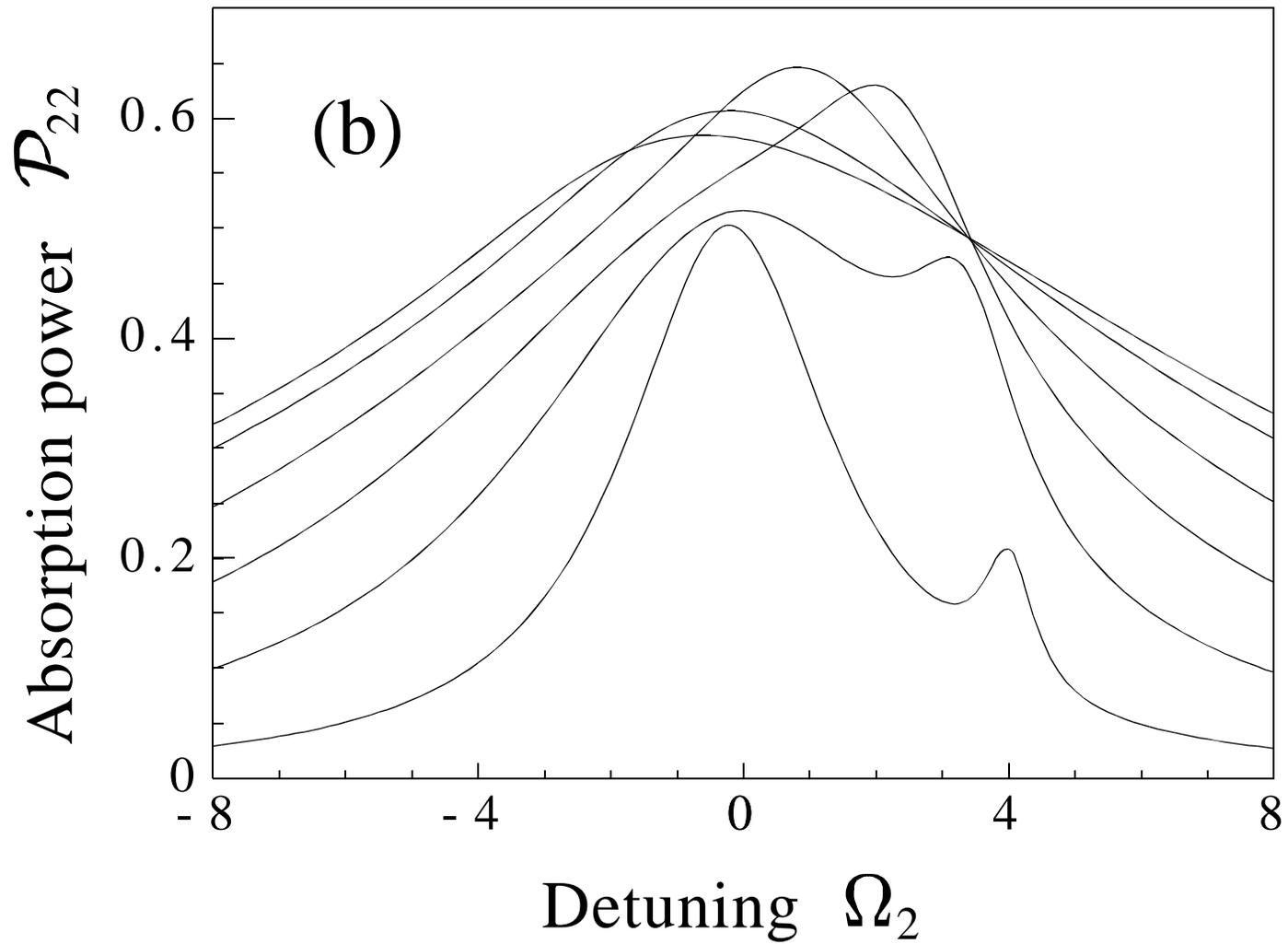
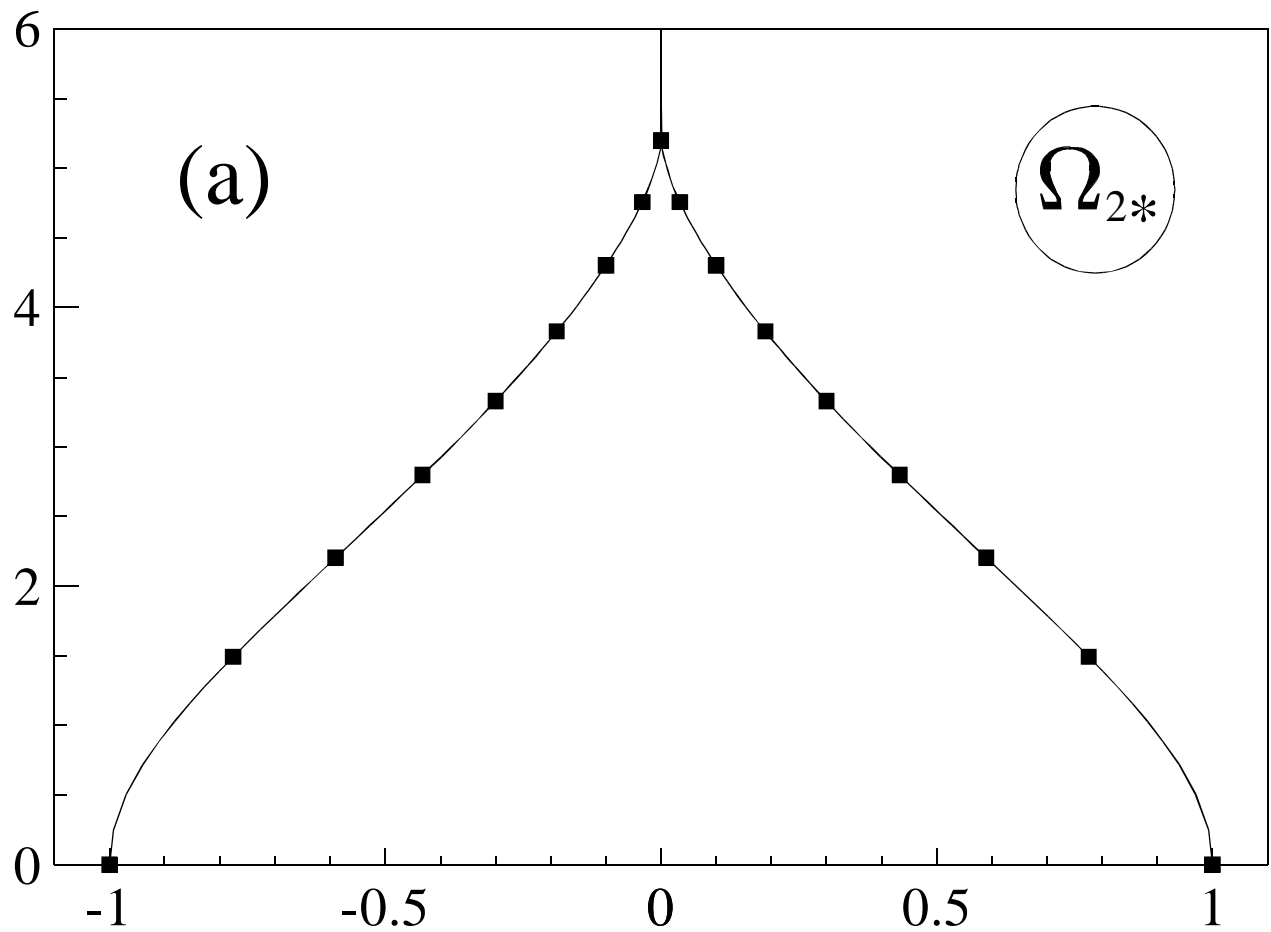
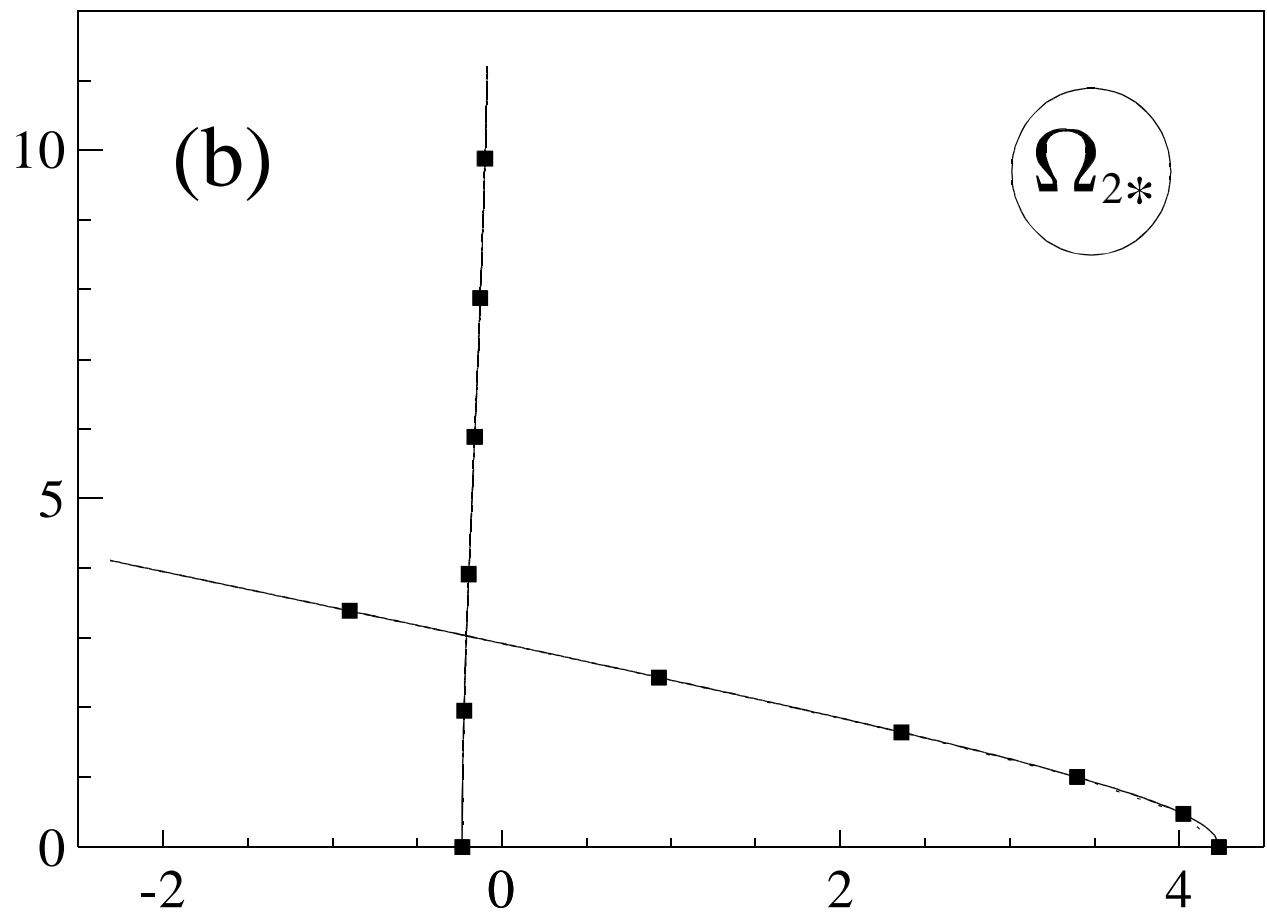
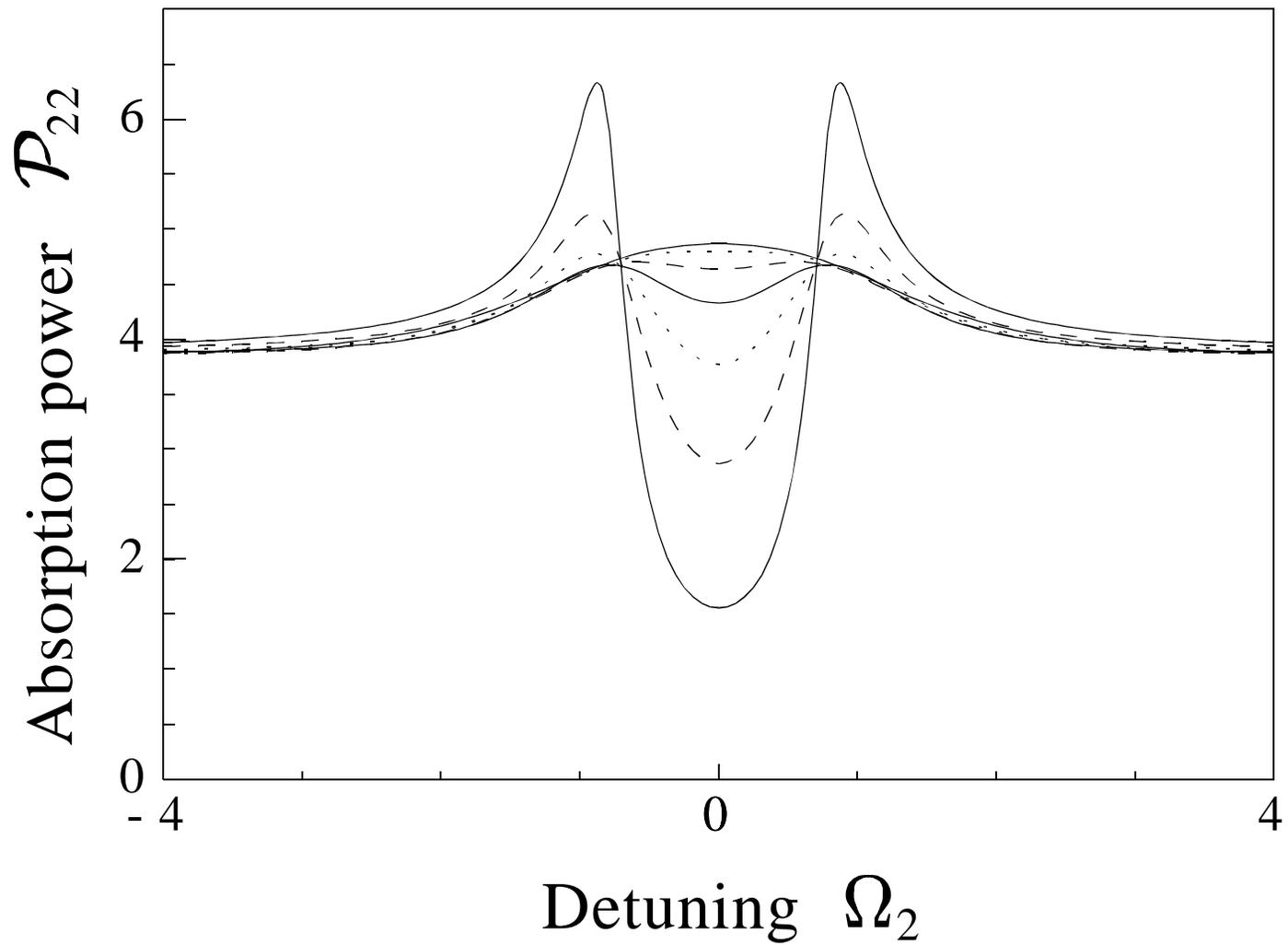
\end{document}